\documentclass[journal]{IEEEtran}

\usepackage{cite}

\usepackage{amsmath,amssymb,amsfonts}
\usepackage{algorithmic}
\usepackage{graphicx}
\usepackage{textcomp}
\usepackage{xcolor}

\newcommand*\diff{\mathop{}\!\mathrm{d}}

\usepackage{url} 

\usepackage{bm}
\usepackage[normalem]{ulem} 
\usepackage{supertabular}
\usepackage{amsthm}
\theoremstyle{definition}

\usepackage[linesnumbered,ruled,vlined]{algorithm2e}
\usepackage{pbox}
\usepackage{float} 
\usepackage{subcaption} 

\begin{document}

\title{Statistical Analysis of Downlink Zero-Forcing Beamforming}

\author{Hussein~A.~Ammar\IEEEauthorrefmark{1},~\IEEEmembership{Student Member,~IEEE}, 
       Raviraj~Adve\IEEEauthorrefmark{1},~\IEEEmembership{Fellow,~IEEE},
       Shahram~Shahbazpanahi\IEEEauthorrefmark{2},~\IEEEmembership{Senior Member,~IEEE},
       and Gary~Boudreau\IEEEauthorrefmark{3},~\IEEEmembership{Senior Member,~IEEE}
       \thanks{
               \IEEEauthorrefmark{1}H. A. Ammar and R. Adve are with the Edward S. Rogers Sr. Department of Electrical and Computer Engineering, University of Toronto, Toronto, ON M5S 3G4, Canada (e-mail: ammarhus@ece.utoronto.ca; rsadve@comm.utoronto.ca).
       }
       \thanks{
               \IEEEauthorrefmark{2}S. Shahbazpanahi is with the Department of Electrical, Computer, and Software Engineering, University of Ontario Institute of Technology, Oshawa, ON L1H 7K4, Canada. He also holds a Status-Only position with the Edward S. Rogers Sr. Department of Electrical and Computer Engineering, University of Toronto.
       }
       \thanks{
               \IEEEauthorrefmark{3}G. Boudreau is with Ericsson Canada, Ottawa, ON L4W 5K4, Canada.
       }
}


\maketitle

\begin{abstract}
We analyze the mean and the variance of the useful signal and interference powers in a multi-cell network using zero-forcing beamforming (ZF-BF) with two beamformer normalization approaches. While the mean has been the main focus in earlier studies on ZF-BF, analysis of the variance has not been tackled. Our analysis provides a complete statistical study, sheds light on the importance of the variance by deriving closed-form expressions for the signals' two moments, and provides a practical use for these expressions; we use the gamma or lognormal distribution for the interference power to analytically calculate the outage.
\end{abstract} 
%




\section{Introduction}\label{sec:introduction}
\IEEEPARstart{M}{\MakeLowercase{ultiple}}-input multiple-output (MIMO) technologies have boosted the capacity, energy efficiency, and performance of wireless communications~\cite{6457363}. 
Achieving the full capabilities of MIMO systems requires serving multiple users on the same time-frequency resource block 
using multiuser beamforming. Zero-forcing beamforming (ZF-BF) is very popular for its analytical tractability and relative ease of implementation. The literature on the use of ZF-BF in MIMO systems is vast, addressing myriad issues. When analyzed, most studies use the properties of Wishart matrices~\cite{1237134} to analyze the effect of metrics such as channel estimations and antenna correlation~\cite{6172680}, receive diversity~\cite{aBackhaul}, and Rician components~\cite{6210404} on the network spectral efficiency. However, crucially, the analysis focuses on studying only the mean value of the studied terms. This, in turn precludes statistical analysis of important metrics that also depend on the second moment of the desired signals and interference. This work aims at filling this~gap.

In this letter, we derive the \emph{two-parameter statistics} - the mean and the variance - of the signal and interference powers received at the users in the network using ZF-BF. To the best of the authors' knowledge, this is the first work to provide accurate closed-form expressions for both of these parameters in a multi-cell network, which uses two different techniques for enforcing the power constraint. To illustrate the potential use of our work, we \emph{derive} the outage probability in the considered network by characterizing the interference power using either the gamma or lognormal distribution. Additionally, it is worth noting that our work is different from that in~\cite{ConjBeamANDZeroForcing6415389} which focuses on accounting for the channel estimation error in a single cell scenario using average normalization for the beamformer to satisfy the power constraint and does not provide a complete statistical analysis. Hence, it serves a different purpose compared to our work.
\section{System Model}
We consider a cellular  MIMO system with $Q$ cells, each containing a base station (BS) which uses ZF-BF to serve $K$ users on the same time-frequency resource block. Each BS is equipped with $M > K$ antennas, while each user is equipped with a single antenna. We consider the effect of Rayleigh small-scale fading and the path loss and neglect shadowing\footnote{We neglect shadowing to minimize notations, but accounting for it is straightforward.}. Moreover, we assume flat fading and perfect channel state information (CSI) at the transmitter.

Let us define the ZF pre-coding matrix as ${\bf W}_{q} \triangleq[{\bf w}_{q1} \dots {\bf w}_{qK}]\in\mathbb{C}^{M\times K}$, where ${\bf w}_{qk} \in\mathbb{C}^{M}$ is the beamforming vector serving user  $k$ in cell $q$. We can then express the signal received at user $k$ in cell $q$ as    
\[ y_{qk} = \mathbf{w}_{qk}^H\mathbf{h}_{q,qk} s_{qk} + \sum_{q' \ne q}\sum_{k'=1}^K \mathbf{w}_{q'k'}^H \mathbf{h}_{q',qk} s_{q'k'} + z_{qk}, \]
where $s_{qk}$ is the data for user $k$ satisfying $\mathbb{E}\{|s_{qk}|^2\}=p$, and $p$ is the transmission power of BS $q$. The vector $\mathbf{h}_{q,qk}$ is the channel between BS $q$ and its $k^\mathrm{th}$ user, $\mathbf{h}_{q',qk}$ denotes the interference channel from cell $q'$ and $z_{qk}$ denotes white Gaussian noise with variance $\sigma_z^2$. Furthermore, we assume ${\bf h}_{q,qk} = \sqrt{\ell(d_{q,qk})} {\bf g}_{q,qk}$, where $\ell(d_{q,qk})$ is the large-scale fading accounting for the path loss and depends on the distance $d_{q,qk}$ between BS $q$ and user $k$ in cell $q$, and ${\bf g}_{q,qk}\sim\mathcal{CN}({\bf 0},{\bf I}_M)$ is the small-scale fading. 

Denoting ${\bf H}_q = \left[{\bf h}_{q,q1}\dots {\bf h}_{q,qK} \right] \in \mathbb{C}^{M \times K}$ as the $M\times K$ channel matrix, we write the precoding matrix in cell $q$ as
\begin{align}\label{eq:beamformerAccess}
{\bf W}_{q}={\bf \widetilde{W}}_{q}\bm{\mu}_q=\left({\bf H}_{q}\right)^\dagger \bm{\mu}_q = {\bf H}_{q}\left({\bf H}_{q}^H{\bf H}_{q}\right)^{-1} \bm{\mu}_q
\end{align}
where we define ${\bf \widetilde{W}}_{q} = {\bf H}_{q}\left({\bf H}_{q}^H{\bf H}_{q}\right)^{-1}$.
Matrix $\bm{\mu}_q \in \mathbb{C}^{K \times K}$ is  diagonal and provides normalizing factors to satisfy the power budget constraint, and it can be designed in one of two cases:
\begin{itemize}
        \item Case 1: we can choose  the $k^{th}$ diagonal entry of $\bm{\mu}_q$ as
        \begin{align}\label{eq:instantaneous_norm}
        \mu_{qk} = \left[\bm{\mu}_q\right]_{k,k} = \sqrt{K^{-1} \|{\bf \widetilde{w}}_{qk}\|^{-2}}
        \end{align}
  which uses the \emph{instantaneous} value of ${\bf \widetilde{w}}_{qk}$ and allows to normalize the instantaneous power as $\text{tr}\left\{ {\bf W}_{q} {\bf W}_{q}^H\right\} = 1$. Such normalization is adopted in works such as~\cite{RequestedByEditor5730587}.     %
        \item Case 2: alternatively, we can choose the $k^{th}$ diagonal entry of $\bm{\mu}_q$ as $\bar{\mu}_{qk}$ given as
        \begin{align}\label{eq:average_norm}
        \bar{\mu}_{qk} = \left[\bm{\mu}_q\right]_{k,k} = \sqrt{K^{-1} \mathbb{E}\left\{\|{\bf \widetilde{w}}_{qk}\|^2 \right\}^{-1}}
        \end{align}
        which ensures that $\mathbb{E}\left\{\text{tr}\left\{ {\bf W}_{q} {\bf W}_{q}^H\right\}\right\} = 1$ holds true. Such normalization is adopted in works such as~\cite{ConjBeamANDZeroForcing6415389}.

\end{itemize}
Note that ${\bf \widetilde{w}}_{qk}$ is the $k^{th}$ column of ${\bf \widetilde{W}}_{q}$, denoted as $\left[ {\bf \widetilde{W}}_{q} \right]_{.k}$.

Both choices, $\mu_{qk}$ and $\bar{\mu}_{qk}$, can be realized in a network, but the analysis for the mean and the variance of the useful signal and interference is different for each choice. This is because, in the first case, the useful signal power $S_{qk}$ is random while in the second it is a constant. 
Moreover, the spectral efficiency~is
\begin{align}
R_{qk}&= 
\log\left(1+\frac{S_{qk}} {I_{qk} + \sigma_z^2} \right)
\hspace*{0.1in} \mathrm{(in~nats/s/Hz)}
\end{align}
where $I_{qk}$ is the inter-cell interference power and $\sigma_z^2$ is the noise power.
%
\section{Normalization based on Instantaneous Power}\label{section:instant}
Using~\eqref{eq:instantaneous_norm}, we ensure that $\text{tr}\left\{ {\bf W}_{q} {\bf W}_{q}^H\right\} = 1$ holds true in each transmission from the BS. In this case the mean of the useful signal power can be derived as
\begingroup
\allowdisplaybreaks
\begin{align}
&\mathbb{E}\left\{S_{qk}\right\} = p \mathbb{E}\left\{{\mu_{qk}}^2 \right\} = \frac{p}{K} \mathbb{E}\left\{\|{\bf \widetilde{w}}_{qk}\|^{-2}  \right\}
\nonumber \\
&
= \frac{p}{K} \mathbb{E}\left\{ \left({\bf \widetilde{w}}_{qk}^H {\bf \widetilde{w}}_{qk}\right)^{-1} \right\}
= \frac{p}{K} \mathbb{E}\left\{ \left( \left[{\bf \widetilde{W}}_{qk}^H {\bf \widetilde{W}}_{qk}\right]_{kk} \right)^{-1} \right\}
\nonumber \\
&
=\frac{p}{K} \mathbb{E}\left\{ \left( \left[\left(\left({\bf H}_{q}^H{\bf H}_{q}\right)^{-1}\right)^H {\bf H}_{q}^H {\bf H}_{q} \left({\bf H}_{q}^H{\bf H}_{q}\right)^{-1}\right]_{kk}\right)^{-1} \right\}
\nonumber \\
&
= \frac{p}{K} \mathbb{E}\left\{ \left(\left[\left({\bf H}_{q}^H{\bf H}_{q}\right)^{-1}\right]_{kk}\right)^{-1} \right\}
\nonumber \\
&
= \frac{p}{K}\ell(d_{q,qk}) \mathbb{E}\left\{ \left(\left[\left({\bf G}_{q}^H{\bf G}_{q}\right)^{-1}\right]_{kk}\right)^{-1} \right\}
\end{align}
The complex variable $X = \left(\left[\left({\bf G}_{q}^H{\bf G}_{q}\right)^{-1}\right]_{kk}\right)^{-1}$ is a scaled chi-square random variable (RV) with a probability density function (PDF) $f(x) = \frac{1}{\left(M - K\right)!} x^{M - K} e^{-x}$~\cite{eaton2007wishart, RequestedByEditor5730587}, i.e., $Y = 2 X$ is a chi-square RV with $2 \left(M - K + 1\right)$ degrees of freedom. Accordingly, following this PDF, $X$ is a gamma distributed RV with shape parameter $(M - K + 1)$ and scale parameter equal to one. This allows us to obtain the mean of $X$ as $(M - K + 1)$. Hence,
\begin{align} \label{eq:S_qn_mu_q_instant}
&\mathbb{E}\left\{S_{qk}\right\} = 
\frac{p \left(M - K + 1\right) \ell(d_{q,qk})}{K}
\end{align}
\endgroup
As for the variance of $S_{qk}$, using the properties of the gamma distribution, we have
\begin{align}
\textbf{Var}&\left\{S_{qk}\right\} = \left( \frac{p}{K} \right)^2 \ell(d_{q,qk})^2 \left(M - K + 1\right)
\end{align}
The interference power, $I_{qk}$, requires a more detailed derivation. The mean can be derived as 
\begingroup
\allowdisplaybreaks
\begin{align}
&\mathbb{E}\left\{I_{qk}\right\} = p\sum_{q' \neq q} \sum_{k'= 1}^{K} \mathbb{E}\left\{ \| {\bf h}_{q',qk}^H {\bf w}_{q'k'} \|^2 \right\}
\nonumber \\
&
= p\sum_{q' \neq q} \sum_{k'= 1}^{K} \mathbb{E}\left\{ {\bf w}_{q'k'}^H {\bf h}_{q',qk} {\bf h}_{q',qk}^H {\bf w}_{q'k'} \right\}
\nonumber \\
&\stackrel{(a)}{=}
p \sum_{q' \neq q} \sum_{k'= 1}^{K} K^{-1} \mathbb{E}\left\{\|{\bf \widetilde{w}}_{q'k'}\|^{-2} \right\} \ell(d_{q',qk})
\nonumber \\
&\quad \times
\mathbb{E}\left\{ {\bf \widetilde{w}}_{q'k'}^H {\bf g}_{q',qk} {\bf g}_{q',qk}^H {\bf \widetilde{w}}_{q'k'}\right\}
\nonumber \\
&\stackrel{(b)}{=} p \sum_{q' \neq q} \sum_{k'= 1}^{K} K^{-1} \mathbb{E}\left\{\|{\bf \widetilde{w}}_{q'k'}\|^{-2} \right\} \ell(d_{q',qk})
\nonumber \\
& \quad \times
\mathbb{E}\left\{ {\bf \widetilde{w}}_{q'k'}^H \mathbb{E}\left\{ {\bf g}_{q',qk} {\bf g}_{q',qk}^H \right\} {\bf \widetilde{w}}_{q'k'} \right\}
\nonumber \\
& =
\resizebox{0.46\textwidth}{!}
{$\displaystyle
        p \sum_{q' \neq q} \sum_{k'= 1}^{K} K^{-1} \mathbb{E}\left\{\|{\bf \widetilde{w}}_{q'k'}\|^{-2} \right\} \ell(d_{q',qk})
        \mathbb{E}\left\{\|{\bf \widetilde{w}}_{q'k'}\|^2 \right\}
        $}
\nonumber \\
& =
\resizebox{0.46\textwidth}{!}
{$\displaystyle
        p \sum_{q' \neq q} \sum_{k'= 1}^{K} K^{-1} \mathbb{E}\left\{\|{\bf \widetilde{w}}_{q'k'}\|^{-2} \right\} \ell(d_{q',qk})
        \mathbb{E}\left\{ \left[{\bf \widetilde{W}}_{q'k'}^H {\bf \widetilde{W}}_{q'k'}\right]_{k'k'} \right\}
        $}
\nonumber \\
& =
\resizebox{0.46\textwidth}{!}
{$\displaystyle
        p \sum_{q' \neq q} \sum_{k'= 1}^{K} K^{-1} \mathbb{E}\left\{\|{\bf \widetilde{w}}_{q'k'}\|^{-2} \right\} \ell(d_{q',qk})
        \mathbb{E}\left\{ \left[\left({\bf H}_{q'}^H{\bf H}_{q'}\right)^{-1}\right]_{k'k'} \right\}
        $}
\nonumber \\
& =
\resizebox{0.47\textwidth}{!}
{$\displaystyle
p \sum_{q' \neq q} \sum_{k'= 1}^{K} K^{-1} \mathbb{E}\left\{\|{\bf \widetilde{w}}_{q'k'}\|^{-2} \right\} \ell(d_{q',qk})
\frac{ \mathbb{E}\left\{
        \text{{\bf tr}}\left\{
        \left({\bf G}_{q'}^H{\bf G}_{q'}\right)^{-1}
        \right\}
        \right\}}{K\ell(d_{q',q'k'})}
$}
\nonumber \\
& \stackrel{(c)}{=}
\resizebox{0.46\textwidth}{!}
{$\displaystyle
p \sum_{q' \neq q} \sum_{k'= 1}^{K} K^{-1} \mathbb{E}\left\{\|{\bf \widetilde{w}}_{q'k'}\|^{-2} \right\} \ell(d_{q',qk})
\frac{\left(M-K\right)^{-1}}{\ell(d_{q',q'k'})}
$}
\nonumber \\
&  \stackrel{(d)}{=}
p \sum_{q' \neq q} \sum_{k'= 1}^{K} K^{-1} \left(\frac{M-K+1}{M-K}\right) \ell(d_{q',qk})
\nonumber \\
& =
p \sum_{q' \neq q} \left(\frac{M-K+1}{M-K}\right) \ell(d_{q',qk})
\end{align}
\endgroup
where $(a)$ follows from the fact that $\bm{\mu}_{q'}$ is a diagonal matrix, hence ${\bf w}_{q'k'} = \left[ {\bf W}_{q'} \right]_{.k'} = {\bf \widetilde{W}}_{q'} \left[ \bm{\mu}_{q'} \right]_{.k'} = \mu_{q'k'} {\bf \widetilde{w}}_{q'k'}$, $(b)$ follows from the independence of ${\bf g}_{q',qk}$ and ${\bf \widetilde{w}}_{q'k'}$, and $(c)$ follows from the fact that ${\bf G}_{q'}^H{\bf G}_{q'} \sim \mathcal{W}_K\left(M,{\bf I}_{M}\right)$ is a $K \times K$ central complex Wishart matrix with $M$ degrees of freedom, hence $\mathbb{E}\left\{\text{{\bf tr}}\left\{\left({\bf G}_{q'}^H{\bf G}_{q'}\right)^{-1}\right\}\right\} = \frac{K}{M - K}$~\cite{1237134}. As for $(d)$, it follows from the results in \eqref{eq:S_qn_mu_q_instant}. At last, for $M - K \gg 1$ the term $\left(\frac{M-K+1}{M-K}\right) \rightarrow 1$. 

As for the variance of the interference power $I_{qk}$, we have
\begin{align}\label{eq:variance_Iqn}
\textbf{Var}\left\{I_{qk}\right\} &=
\resizebox{0.39\textwidth}{!}
{$\displaystyle
        \sum_{q' \neq q} \sum_{k'= 1}^{K} \left(p \ell(d_{q',qk})\right)^2 \textbf{Var}\left\{\| {\bf g}_{q',qk}^H {\bf w}_{q'k'} \|^2\right\}
        $}
\nonumber \\
&
\simeq
\frac{1}{K}
\sum_{q' \neq q} \left(p \ell(d_{q',qk})\right)^2
\end{align}
where the last step follows from treating ${\bf g}_{q',qk}^H {\bf w}_{q'k'}$ as complex Gaussian RVs, and hence the norm $X=\| {\bf g}_{q',qk}^H {\bf w}_{q'k'} \|^2$ has an exponential distribution of rate $K$, i.e., its variance is $1/K^2$ leading to $\sum_{k'= 1}^{K} 1/K^2 = 1/K$ as shown in~\eqref{eq:variance_Iqn}.
%
\section{Normalization based on Average Power}
\par Using $\bar{\mu}_{qk}$ as a normalizing term for the beamformer produces $\mathbb{E}\left\{\text{tr}\left\{ {\bf W}_{q} {\bf W}_{q}^H\right\}\right\} = 1$, and this results in the useful signal power being a constant. Hence, the variance of the useful signal power is zero, while its mean is calculated as
\begingroup
\allowdisplaybreaks
\begin{align} \label{eq:S_qn_mu_q_average}
S_{qk}&
= p \bar{\mu}_{qk}^2 = \frac{p}{K} \mathbb{E}\left\{\|{\bf \widetilde{w}}_{qk}\|^2 \right\}^{-1}
\nonumber \\
&
= \frac{p}{K} \mathbb{E}\left\{{\bf \widetilde{w}}_{qk}^H {\bf \widetilde{w}}_{qk} \right\}^{-1}
= \frac{p}{K} \mathbb{E}\left\{ \left[{\bf \widetilde{W}}_{qk}^H {\bf \widetilde{W}}_{qk}\right]_{kk} \right\}^{-1}
\nonumber \\
&
=\frac{p}{K} \mathbb{E}\left\{ \left[\left(\left({\bf H}_{q}^H{\bf H}_{q}\right)^{-1}\right)^H {\bf H}_{q}^H {\bf H}_{q} \left({\bf H}_{q}^H{\bf H}_{q}\right)^{-1}\right]_{kk} \right\}^{-1}
\nonumber \\
&
= \frac{p}{K} \mathbb{E}\left\{ \left[\left({\bf H}_{q}^H{\bf H}_{q}\right)^{-1}\right]_{kk} \right\}^{-1}
\nonumber \\
&
\stackrel{(a)}{=} \frac{p \left(M - K\right) \ell(d_{q,qk})}{K}
\end{align}
\endgroup
where $(a)$ follows since ${\bf G}_{q}^H{\bf G}_{q}$ is a central Wishart matrix as noted earlier, an approach also used by others, e.g.,~\cite{1237134, 6457363}.

Both analyses in \eqref{eq:S_qn_mu_q_instant} and \eqref{eq:S_qn_mu_q_average} give approximately the same mean for the useful signal power received at the user, especially when $M$ is large enough compared to $K$. But the mean of $S_{qk}$ when using $\mu_{qk}$ should be always larger because $1/X$ is convex for any strictly positive $X$, hence we have $\mathbb{E}\{1/X\} \ge 1/\mathbb{E}\{X\}$. Therefore, $\mathbb{E}\left\{\|{\bf \widetilde{w}}_{qk}\|^{-2}  \right\} \ge \mathbb{E}\left\{\|{\bf \widetilde{w}}_{qk}\|^2 \right\}^{-1}$ is always true.

As for $I_{qk}$, it can be analyzed in a similar fashion as
\begingroup
\allowdisplaybreaks
\begin{align}
&\mathbb{E}\left\{I_{qk}\right\} = p\sum_{q' \neq q} \sum_{k'= 1}^{K} \mathbb{E}\left\{ \| {\bf h}_{q',qk}^H {\bf w}_{q'k'} \|^2 \right\}
\nonumber \\
&
= p\sum_{q' \neq q} \sum_{k'= 1}^{K} \mathbb{E}\left\{ {\bf w}_{q'k'}^H {\bf h}_{q',qk} {\bf h}_{q',qk}^H {\bf w}_{q'k'} \right\}
\nonumber \\
&\stackrel{(a)}{=}
\resizebox{0.46\textwidth}{!}
{$\displaystyle
p \sum_{q' \neq q} \sum_{k'= 1}^{K} K^{-1}\mathbb{E}\left\{\|{\bf \widetilde{w}}_{q'k'}\|^2 \right\}^{-1}
\mathbb{E}\left\{ {\bf \widetilde{w}}_{q'k'}^H {\bf h}_{q',qk} {\bf h}_{q',qk}^H {\bf \widetilde{w}}_{q'k'}\right\}
$}
\nonumber \\
& =  p \sum_{q' \neq q} \sum_{k'= 1}^{K} K^{-1} \mathbb{E}\left\{\|{\bf \widetilde{w}}_{q'k'}\|^2 \right\}^{-1} \ell(d_{q',qk})
\nonumber \\
&\quad \times
\mathbb{E}\left\{ {\bf \widetilde{w}}_{q'k'}^H {\bf g}_{q',qk} {\bf g}_{q',qk}^H {\bf \widetilde{w}}_{q'k'}\right\}
\nonumber \\
&= p \sum_{q' \neq q} \sum_{k'= 1}^{K} K^{-1} \mathbb{E}\left\{\|{\bf \widetilde{w}}_{q'k'}\|^2 \right\}^{-1} \ell(d_{q',qk})
\nonumber \\
& \quad \times
\mathbb{E}\left\{ {\bf \widetilde{w}}_{q'k'}^H \mathbb{E}\left\{ {\bf g}_{q',qk} {\bf g}_{q',qk}^H \right\} {\bf \widetilde{w}}_{q'k'} \right\}
\nonumber \\
& =
\resizebox{0.46\textwidth}{!}
{$\displaystyle
p \sum_{q' \neq q} \sum_{k'= 1}^{K} K^{-1} \mathbb{E}\left\{\|{\bf \widetilde{w}}_{q'k'}\|^2 \right\}^{-1} \ell(d_{q',qk})
\mathbb{E}\left\{\|{\bf \widetilde{w}}_{q'k'}\|^2 \right\}
$}
\nonumber \\
& = p \sum_{q' \neq q} \sum_{k'= 1}^{K} \ell(d_{q',qk})K^{-1} 
= p\sum_{q' \neq q} \ell(d_{q',qk})
\end{align}
\endgroup
where $(a)$ follows from ${\bf w}_{q'k'} = \mu_{q'k'} {\bf \widetilde{w}}_{q'k'}$. As for the variance it is the same as in~\eqref{eq:variance_Iqn}.
\begin{figure}
        \centering
        \includegraphics[width=0.45\textwidth]{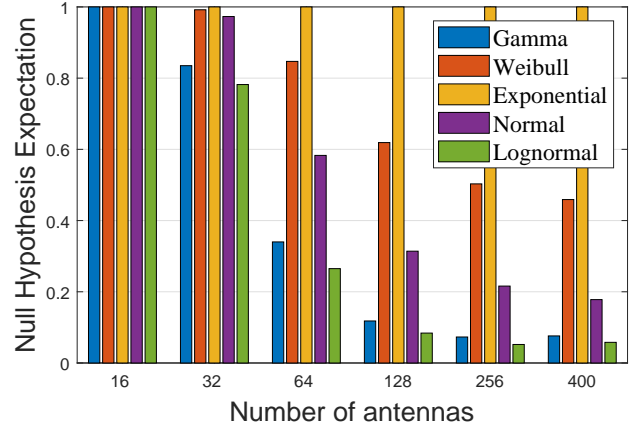}
        \caption{Expectation of KS Tests over simulation trials for the interference signal when using $\mu_{qk}$.}
        \label{fig:KSTest}
\end{figure}

\section{Characterizing Outage via Moment Matching} \label{sec:Outage}
%
%
%
%
%
\begin{figure*}[t]
        \vspace{-1em}
        \centering
        \begin{subfigure}[t]{0.49\textwidth}
                \centering
                \includegraphics[width=1\textwidth]{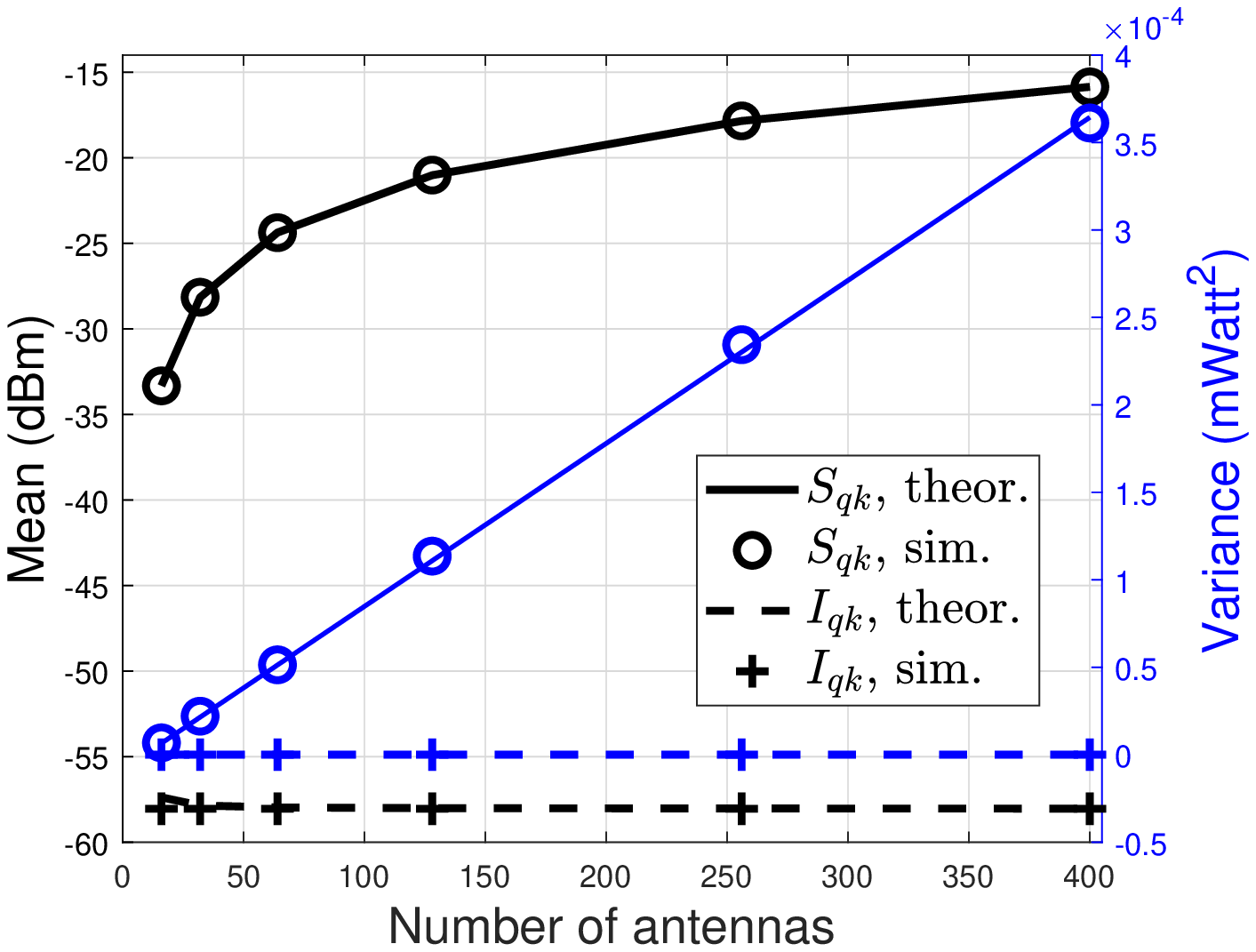}
                \label{sfig:MeanVar_inst}
        \end{subfigure}
        $\ $
        \begin{subfigure}[t]{0.49\textwidth}
                \centering
                \includegraphics[width=0.95\textwidth]{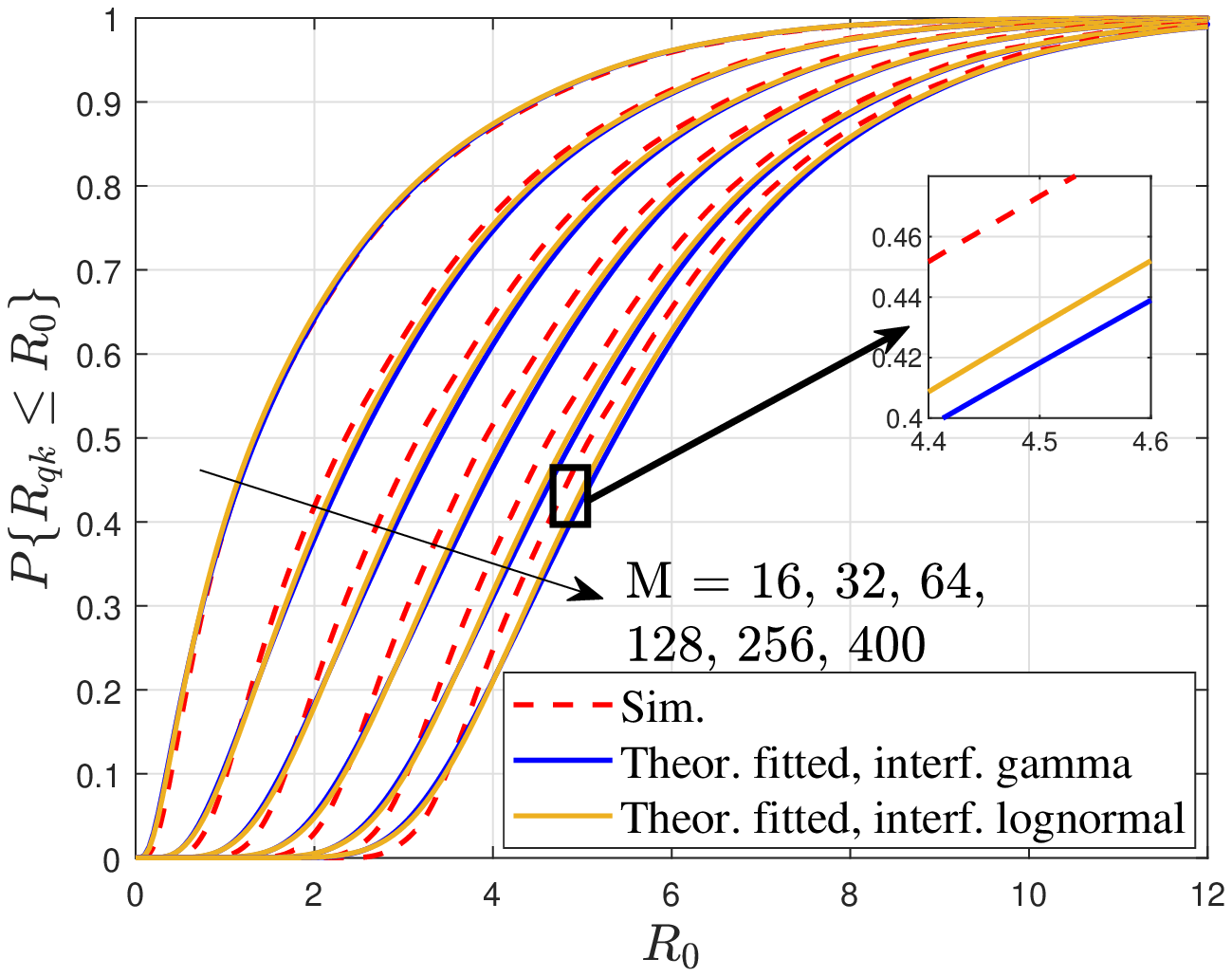}
                \label{sfig:Outage_inst}
        \end{subfigure}
        \caption{Case 1: using $\mu_{qk}$; (a) Mean and variance accuracy (legend applies to black and blue curves), (b) Obtained outage.}
        \label{fig:mu_qn_ins}
\end{figure*}
\begin{figure*}[t]
        \vspace{-1em}
        \centering
        \begin{subfigure}[t]{0.49\textwidth}
                \centering
                \includegraphics[width=1\textwidth]{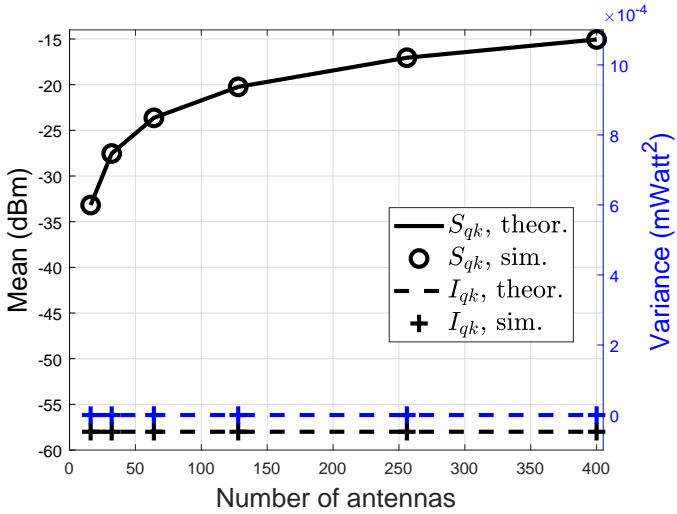}
                \label{sfig:meanvar_ZFBF_mu_qn}
        \end{subfigure}
        $\ $
        \begin{subfigure}[t]{0.49\textwidth}
                \centering
                \includegraphics[width=0.95\textwidth]{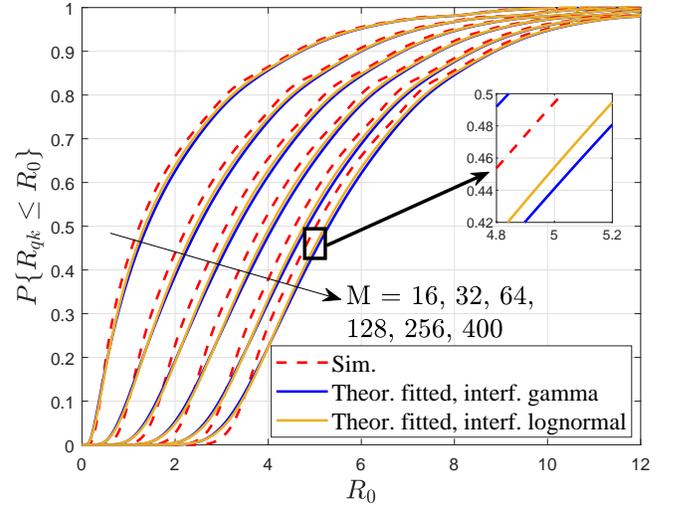}
                \label{sfig:outage_ZFBF_mu_qn}
        \end{subfigure}
        \caption{Case 2: using $\bar{\mu}_{qk}$; (a) Mean and variance accuracy (legend applies to black and blue curves), (b) Obtained outage.}
        \label{fig:mu_qn_avg}
\end{figure*}

%
In this section, we present an important application of our analysis, which is obtaining the outage of ZF-BF transmissions. Theoretically characterizing the outage is very useful when studying network performance. For example, it can be used in optimization frameworks that require imposing a constraint on the outage on the access or on the backhaul channel, e.g.,~\cite{6891348}. For a target rate, $R_0$, the outage in each cell $q$ is given~by
\begin{align}
&P_\text{o} = \mathbb{P}\left\{ R_{qk} \le R_0 \right\} \label{eq:outage}
\\
&\ 
=
\begin{cases}
\int_{0}^{\infty}
F_{S_{qk}}\left(\left(e^{R_0} - 1\right) \left(i + \sigma_z^2\right)\right) f_{I_{qk}}\left(i\right)
\diff i, & \text{using}\ \mu_{qk}\\
1 - F_{I_{qk}}\left(
\frac{ S_{qk} }{e^{R_0} - 1} - \sigma_z^2
\right),  & \text{using}\ \bar{\mu}_{qk}
\end{cases}   \nonumber 
\end{align}
where $F_{S_{qk}}\left(s\right)$ is the cumulative distribution function (CDF) of the signal power and $f_{I_{qk}}\left(i\right)$ is the PDF of the interference power received at user~$k$.
\begin{proof}
        The first equation arises from conditioning on the interference power and then finding the probability of insufficient signal power. In the second, the useful signal is a constant, and hence, outage occurs when the interference is above a threshold. Please check the appendix for more~details.
\end{proof}

To be able to characterize the outage as shown in equation~\eqref{eq:outage}, we need to find an appropriate distribution that characterizes the interference term $I_{qk}$. To do this, we use the Kolmogorov–Smirnov (KS) test, which compares an empirical statistic with a reference distribution to reject or accept the null hypothesis that the compared sample was drawn from the distribution, hence estimating the goodness of the fit. The KS test derives a KS statistic based on the supremum of the distance between the empirical CDF of the data and that of the reference distribution. If the sample comes from reference distribution, the KS statistic converges to zero almost surely as the number of available samples tends to infinity. Then, the null hypothesis is rejected based on the p-value of the KS~statistic.

In Fig.~\ref{fig:KSTest}, we plot the KS test resulting from Monte Carlo simulation in the network setup described in Section~\ref{section:results}. We perform the KS test on the calculated interference power data obtained from each Monte Carlo simulation, and we plot the average result of the null hypothesis. We use a significance level of $5\%$ to reject the hypothesis, where a smaller significance level indicates a more likelihood for the hypothesis to be true. The results show that the normal and the lognormal distributions provide a good fit at higher number of antennas at the BSs, with slightly better results for the lognormal. In general, the studied interference power is small which makes it very sensitive to the small changes in the simulation and hence affects rejecting the null hypothesis. Additionally, we speculate that the lognormal distribution is a better fit for the interference because of its heavier right tail property~\cite{cho2004comparison}. As for the useful signal, upon using the $\mu_{qk}$ normalization, it does follow a gamma distribution as stated in Section~\ref{section:instant}.

Equipped with the mean and the variance of the interference power, we can analytically approximate its PDF by a reference distribution. The gamma distribution is characterized by the shape ($K_y$) and scale ($\lambda_y$) parameters of the RV $Y$, and in their turn they are related to the mean and the variance of $Y$~as 
\begin{align} \label{eq:shape_scale_GammaDist}
\resizebox{0.43\textwidth}{!}
{$\displaystyle
        K_y = \frac{\left(\mathbb{E}\{y\}\right)^2}{\textbf{Var}\{y\}} > 0, \ \lambda_y = \frac{\textbf{Var}\{y\}}{\mathbb{E}\{y\}} > 0,
        \ \text{for}\ y \in \{s,i\}
        $}
\end{align}
The gamma distribution has been used with \emph{numerical} fitting in many works e.g.,~\cite{7898403}, where the variance was not analyzed in any detail. Similar expressions can be obtained for the lognormal distribution.

\section{Results}\label{section:results}
To validate our analysis, we consider a network of $Q=9$ square cells (assuming wraparound) of area $1~\text{km}^2$ with a BS at each cell center. We assume that each BS has a power budget of $p = 45$~dBm, and the noise power is $-174$ dBm with a system bandwidth of $900$~kHz corresponding to $5$ resource blocks of $180$~kHz each. Additionally, we assume $K = 10$ users which are uniformly distributed inside each cell with a circular exclusion region with radius $20$~meters around each BS. For the path loss, we use the COST231 Walfish-Ikegami model~\cite{Walfisch14401,1143419} which gives $\ell(d_{q,qk}) = \left(d_{q,qk}/d_0\right)^{-\alpha}$ with a reference distance $d_0 =  1.1$~meters and a path loss exponent $\alpha = 3.8$. These parameters are suitable for a typical cellular network operating at a frequency $1800$~MHz. We validate all of our results using Monte Carlo simulations of $200$ realizations for user locations, each averaged over $1000$ small-scale fading channel realizations.

In Figures~\ref{fig:mu_qn_ins}(a) and~\ref{fig:mu_qn_avg}(a), we plot the mean and the variance of the signal power $S_{qk}$ and interference power $I_{qk}$. As can be seen from these figures, the formulas are very accurate for both small and large number of antennas $M$ on each BS. Additionally, in Figures~\ref{fig:mu_qn_avg}(b) and~\ref{fig:mu_qn_ins}(b), we plot the resulted outage when we use these derived moments as analyzed in Section~\ref{sec:Outage}, where the gamma and lognormal distributions are used to approximate the interference, while the gamma distribution is always used for the useful signal power. The error from approximating $I_{qk}$ as a gamma or lognormal is very small. In this regard, the Root Mean Square Error (RMSE) of the fitted results in Fig.~\ref{fig:mu_qn_ins}(b) ranges from $0.0053$ to $0.0214$ for the different antenna configuration. As for the case of using $\bar{\mu}_{qk}$, i.e., beamformer normalization based on average power, the RMSE ranges from $0.0147$ to $0.0225$ for the fitted results in Fig.~\ref{fig:mu_qn_avg}(b).

The results show that the error from approximating the interference as a gamma or lognormal increases with $M$ due to the larger dimension of the channels and the approximation in~\eqref{eq:variance_Iqn}. In particular, this can be further confirmed by Figures~\ref{fig:mu_qn_ins}(a) and~\ref{fig:mu_qn_avg}(a), which show high accuracy of the formulas for both small and large values of $M$. Nonetheless, the error in the outage is still less than $6\%$ for large $M$, and it is negligible when $M$ is small. Note that using the distribution approximation provides an easy analytical method and allows characterizing the outage with good accuracy.

\section{Conclusion}
We have derived the first two moments for the powers of the useful signal and the interference in a multi-cell network using ZF-BF, which, in turn, allows for statistically characterizing them. We have used two different normalization techniques for the beamformer to satisfy the power budget; the first normalizes the instantaneous power, while the second satisfies the average power. Moreover, we have shown one important application for using these two moments, where we derived the network outage. However, we believe a statistical characterization of ZF-BF will have other applications.

\appendix  
For the instantaneous normalization (Case 1), we have
\allowdisplaybreaks{
\begin{align}\label{eq:appendix}
\mathbb{P}&\left\{R_{qk} \le R_0\right\}
=
\mathbb{P}\left\{\log\left( 1 + \frac{S_{qk}}{I_{qk} + \sigma_z^2} \right) \le R_0
\right\}
\nonumber \\
&
=
\resizebox{0.18\textwidth}{!}
{$\displaystyle
        \mathbb{P}\left\{\frac{S_{qk}}{I_{qk} + \sigma_z^2} \le e^{R_0} - 1
        \right\}
        $}
=
\resizebox{0.24\textwidth}{!}
{$\displaystyle
        \mathbb{P}\left\{S_{qk} \le \left(e^{R_0} - 1\right) \left(I_{qk} + \sigma_z^2\right)
        \right\}
        $}
\nonumber\\
&=
\int_{0}^{\infty} \mathbb{P}\left\{S_{qk} \le \left(e^{R_0} - 1\right) \left(I_{qk} + \sigma_z^2\right)
\right\} f_{I_{qn}}\left(i\right)
\diff i
\nonumber \\
&
=
\int_{0}^{\infty} \int_{0}^{\left(e^{R_0} - 1\right) \left(I_{qk} + \sigma_z^2\right)}
f_{S_{qk}}\left(s\right) f_{I_{qk}}\left(i\right)
\diff s
\diff i
\nonumber\\
& = \int_{0}^{\infty}
F_{S_{qk}}\left(\left(e^{R_0} - 1\right) \left(i + \sigma_z^2\right)\right) f_{I_{qk}}\left(i\right)
\diff i
\end{align}
}
which completes the proof; Case 2 with the average normalization can be solved similarly.
%






\ifCLASSOPTIONcaptionsoff
  \newpage
\fi

\footnotesize
\bibliography{ZFBFLetter_Ref}
\bibliographystyle{unsrt}


\end{document}